\documentstyle[prb,aps]{revtex}
\twocolumn

\begin{document}
\draft
\title{Unusual Field-Dependence of the Intragrain Superconductive Transition in RuSr%
$_{2}$EuCu$_{2}$O$_{8}$}
\author{B. Lorenz$^1$, Y. Y. Xue$^1$, R. L. Meng$^1$, and C. W. Chu$^{1,2}$}
\address{$^1$Texas Center for Superconductivity and Department of Physics, University 
\\
of Houston,\\
Houston, TX 77204-5932, USA}
\address{$^2$Lawrence Berkeley National Laboratory, 1 Cyclotron Road, Berkeley, CA \\
94720, USA}
\date{\today}
\maketitle

\begin{abstract}
A narrow intragrain phase-lock transition was observed in RuSr$_2$EuCu$_2$O$%
_8 $ under a magnetic field H up to a few Tesla. The corresponding
transition temperature, T$_2$, decreases rapidly ($\approx$ 100 K/T at low
fields) with H indicating that the grains of RuSr$_2$EuCu$_2$O$_8$ behave
like a Josephson-junction-array instead of a homogeneous bulk
superconductor. Our data suggest that the bulk superconducting transition
may occur on a length scale well below the grain size of 2 to 6 $\mu m$.
\end{abstract}






\section{Introduction}

Interest has been raised since the reported coexistence of superconductivity
(SC)\cite{1,2}, weak ferromagnetism (FM)\cite{1}, and antiferromagnetism
(AFM)\cite{3,4} in RuSr$_{2}$GdCu$_{2}$O$_{8}$ (Ru1212Gd) and RuSr$_{2}$EuCu$%
_{2}$O$_{8}$ (Ru1212Eu). The superconducting transition in these compounds
show the typical behavior of a granular superconductor. Two steps of the
transition, ascribed to intergrain and intragrain transitions, are well
separated in magnetic susceptibility and electrical resistivity experiments 
\cite{5,6}. However, recent neutron scattering experiments have shown a more
complex nature of the magnetic ordering including a change of the principal
axis of the magnetic moments with an applied external field.\cite{3}
Therefore, the homogeneity of the intragrain magnetic and superconducting
states is questioned. Extrinsic structural defects or the possible phase
separation of FM and AFM regions are expected to have a major effect on the
intragrain superconductivity. In such a scenario a grain of a ceramic Ru1212
sample could actually consist of nanoscale SC domains coupled through
Josephson junctions below a thermodynamic transition temperature, $T_{c}$.
Therefore, the behavior of the intragrain SC under magnetic field, H, is of
particular interest. The magneto-transport properties of Ru1212 should
strongly depend on the intragrain Josephson coupling strength. In the strong
coupling limit, the intragrain Josephson-junction penetration depth ($%
\lambda _{2}$) may roughly be equal to the London penetration depth ($%
\lambda _{L}$) below an intragrain phase-lock transition of $T_{2}$, and the
intragrain SC properties are similar to those of other ''ordinary''
cuprates. In the weak coupling limit, however, individual grains may rather
be similar to disordered Josephson-junction arrays (JJA) with an unusually
large $\lambda _{2}$\cite{8}. In this case a severe suppression of $T_{2}$
by a comparatively small magnetic field is expected.

Another interesting observation is the recently reported field dependence of
a specific heat anomaly near 46 K in Ru1212Gd.\cite{5} It was suggested that
the $T_{c}$ increases with magnetic field $H$ up to 4 to 5 T due to a
possible p-wave superconductivity. Such an abnormal $H$-dependence of $T_{c}$
should show up also in the magneto-transport properties. In the strong
(intragrain) Josephson-coupling limit $T_{2}$ is expected to show a similar
field dependence as $T_{c}$ whereas in the weak-coupling limit $T_{2}$
should decrease with $H$ as is qualitatively known for Josephson junction
arrays. Previous magneto-resistance (MR) measurements below the
ferromagnetic transition temperature, $T_{F}$, show a positive MR at low
fields passing through a maximum at about 2 Tesla but no effect on the
superconducting $T_{c}$ was reported.\cite{9}

Despite the intensive investigations of the last decade the SC properties of
a disordered 2D JJA under magnetic field is still a matter of debate.
Depending on the nature of the disorder and the model used different ground
states have been proposed, e.g. various glass states,\cite{10} metal,\cite
{11} superconductors with global phase coherence,\cite{12} paramagnetic
Meissner state,\cite{13} and chirally ordered normal phase.\cite{14} Various
phase transitions may appear (e.g. glass transitions). The intragrain
magnetotransport of Ru1212, therefore, offers the opportunity to explore 2D
JJA if the coupling strength is moderate to weak, i.e. if the intragrain $%
T_{2}$ is clearly smaller than $T_{c}$ and the penetration depth, $\lambda
_{2}$, is unusually large.

This motivated our investigation of the magneto-transport of Ru1212 ceramic
samples. In order to avoid any additional magnetic contribution from the
Gd-moment the measurements were mainly done on Ru1212Eu which has been shown
to be very similar to Ru1212Gd with respect to the FM and SC transitions. 
\cite{6} AC susceptibility $\chi $ under a $dc$ bias field $H$ combined with
bulk resistivity was used to extract the intragrain SC transition. Various
powder samples were also used to identify the characteristic length scales
involved.

\section{Experimental Setup}

Ceramic samples with a nominal composition RuSr$_{2}$EuCu$_{2}$O$_{8}$ were
prepared by solid-state reaction techniques as described earlier.\cite{6}
The X-ray diffraction pattern shows that the sample is nearly single phase
with a small amount of SrRuO$_{3}$. Elemental analysis reveals a slight Ru
deficiency and some extra copper with the actual cation composition of
Ru:Sr:Eu:Cu=0.91(5):2:1.06(2):2.29(2). From structural, magnetic and
transport properties it was concluded that RuSr$_{2}$EuCu$_{2}$O$_{8}$ is
very similar to its extensively investigated sister compound, RuSr$_{2}$GdCu$%
_{2}$O$_{8}$. In particular, the superconducting transition proceeds in two
steps indicating intergrain and intragrain transitions of a granular
superconductor.\cite{6}

DC and AC susceptibility measurements were perfomed using the Quantum Design
SQUID magnetometer. The AC measurements were done with a field amplitude of
3 Oe in a DC bias field up to 5 Tesla. Magneto-resistance data were measured
in the same SQUID by an AC resistance bridge from Linear Research.

\section{Results and Discussion}

The zero-field cooled (ZFC) and the field cooled (FC) magnetizations of a
bulk piece of Ru1212Eu at 5 Oe are shown in Fig. 1. The magnetic transition
temperature $T_{F}$=115 K is slightly lower than the 133 K observed in
Ru1212Gd probably due to the possible Cu/Ru site mixing as suggested by the
cation nonstoichiometry. The intergrain coupling seems to be rather weak so
that the nonlinearity of the ZFC $M-H$ branch begins below 1 Oe and $M$
saturates at 4 Oe (Inset, Fig. 1). Two drops of $M_{ZFC}$ appear at 20 K and
28 K, respectively. The susceptibility ($M_{ZFC}/H$) decreases quickly below
20 K and reaches -0.04 emu/cm$^{3}$ (about 50\% shielded-volume fraction) at
5 K. The magnetization of a powder sample with an average particle size of
about 3 $\mu $m is also shown in Fig. 1 (solid lines). Based on the
comparison of the data we conclude that the 20 K transition indicates the
intergrain phase-lock temperature and the 28 K diamagnetic transition is the
intragrain transition at $T_{2}$. This interpretation is further supported
by the AC susceptibility data shown in Fig. 2. The peaks in the imaginary
part, $\chi "$, and the drops in the real part, $\chi ^{\prime }$, appear
around 20 K and 30 K, respectively. Note that these temperatures are about
10 K lower than those of Ru1212Gd ($T_{1}\approx $30 K and $T_{2}\approx $40
K),\cite{8} probably due to the chemical pressure caused by the replacement
of Gd by the smaller Eu. The diamagnetic drop between $T_{1}$ and $T_{2}$, $%
\Delta $M$_{ZFC}\approx $0.002 emu/cm$^{3}$, in Ru1212Eu is significantly
larger than that in Ru1212Gd\cite{2,8,15} suggesting a shorter $\lambda _{2}$
in our Ru1212Eu sample. The larger intragrain Meissner effect, $\Delta M_{FC}
$, above $T_{1}$ observed here seems to be closely related to the shorter $%
\lambda _{2}$ but will not be discussed further since it is not essential to
the topic presented here.

The zero field resistivity, $\rho (0)$, and thermoelectric power, S, of the
bulk piece of Ru1212Eu have been discussed previously.\cite{6} $\rho $
reaches a maximum around $T_{m}=36K$ and drops to zero slightly above 20 K.
The thermoelectric power changes slop close to $T_{m}$ and decreases to zero
slightly above the zero resistance temperature. The details of the resistive
transition are shown in the derivative, d$\rho $/d$T$ (Fig. 3). In zero
magnetic field (Fig. 3a) d$\rho $/d$T$ appears as a close superposition of
two peaks corresponding to the inter- and intragrain superconducting
transitions. The two peaks are split off by small magnetic field and can
easily decomposed by assuming Gaussian peak shapes (dotted lines in Fig.
3b). This procedure is used to separate the contributions from both
transitions to the overall drop of resistivity and to estimate the
transition temperatures, $T_{1}$ and $T_{2}$, respectively. In particular,
the deconvolution of the two peaks allows us to determin the temperature $%
T_{2}$ where the intragrain resistivity is close to zero. This temperature
should roughly correspond to the $T_{2}$ identified in the magnetization
measurements. In the following evaluation we define $T_{2}$ from resistivity
by a 95 \% drop of the intragrain resistance as indicated by the double
sided arrow in Fig. 3b. The chosen criterion may appear somewhat arbitrary
but it does not affect the main conclusions.\ The maximum of resistivity
appears at higher temperature, $T_{m}$, defined by th zero crossing of d$%
\rho $/d$T$.

The intergrain transition temperature, $T_{1}$, decreases quickly in an
external magnetic field. This is typical for a granular superconductor with
weak intergrain coupling. The field dependence of the intragrain
temperature, $T_{2}$, is of special interest and is estimated from both,
magneto-resistivity data as described above, and AC susceptibility
experiments in a DC bias field. A typical AC sysceptibility curve is shown
in Fig. 4 for an external DC field of 1000 Oe. The intragrain transition is
clearly detected by the drop of susceptibility. $T_{2}$ is determined from
the crossing of the two extrapolated linear parts of $\chi $ above and below
the transition (Fig. 4). Surprisingly, $T_{2}$ is rapidly suppressed by
small magnetic fields with an initial slope of 100 K/T (Fig. 5). The $T_{2}$%
's estimated from resistivity and susceptibility are shown in Fig. 5 by open
triangles and closed circles, respectively. The agreement of both data sets
is very good, the deviation at $H\approx 0$ is obviously due to the larger
uncertainty in the deconvolution of the two peaks in d$\rho $/d$T$ at very
small field (Fig. 3a). The unusually strong decrease of the intragrain $T_{2}
$ of d$T_{2}/$d$H\approx $100\ K/T at low field cannot be explained by bulk
and homogeneous superconductivity inside the grains. However, a steep field
dependence of $T_{2}$ is expected if the intragrain transition is considered
to be a phase-lock transition of another (intragrain) Josephson junction
array. The underlying subgrain structures could be structural or magnetic
domains of typical nanometer size. The broad intragrain resistivity
transition, in particular the broadening of the d$\rho $/d$T$ peak with
field $H$, might be understood as a percolative transition between domains
coupled by junctions of different strength and disorders induced by $H$. The
AC susceptibility, however, shows a narrow phase-lock transition (width $%
\approx $ 0.7 K) between the two linear sections (Fig. 4) indicating a true
phase-transition. The width of this transition does not change with magnetic
field. The linear decrease of $\chi $ below $T_{2}$ can be explained by the
decrease of the penetration depth, $\lambda _{2}$. From an analysis of the
intragrain superconducting transitions of powders with different particle
size it is concluded that the penetration depth of the Ru1212Eu sample used
here is $\lambda _{2}(0K)\approx 1\ \mu m$. Details of this analysis will be
published elsewhere.

Besides $T_{1}$ and $T_{2}$, the maximum in resistivity ($T_{m}$) and the
resistance in the ferromagnetic phase, $\rho (H)$, are relevant quantities
for investigating the interplay of superconductivity and magnetic
structures. At low field, up to about 0.4 T, $\rho (H)$ and $T_{m}$ are not
changed by the magnetic field. For $H>0.4$ T, however, the resistivity
increases and its maximum shifts to higher temperature, saturating at about
4 to 5 T. As shown in Fig. 6, $\rho (H)$ above 50 K and $T_{m}(H)$ seem to
be strongly correlated. A positive magneto-resistance similar to the present
data for Ru1212Eu has also been reported for Ru1212Gd below the
ferromagnetic transition temperature.\cite{9} The anomalous increase of $%
T_{m}(H)$ and $\rho (H)$ above 0.4 T may be due to a change in the magnetic
structure of Ru1212Eu as previously proposed from neutron scattering
experiments on Ru1212Gd.\cite{3} According to the neutron scattering data
the Ru moments are ordered antiferromagnetically with a small residual
ferromagnetic moment. The direction of the moments was suggested to be
parallel to the tetragonal c-axis. At 0.4 T the Ru moments are assumed to
rotate into another antiferromagnetic structure and a sizable ferromagnetic
magnetization can be induced at higher fields. This change in the magnetic
structure is obviously reflected in the positive magneto-resistivity and the
increase of $T_{m}$ as discussed above. As a consequence $\rho (H)$ and $%
T_{m}(H)$ should be correlated. In fact, the inset in Fig. 6 indicates a
linear relation between $\rho $ and $T_{m}$, but a change of slope possibly
takes place at a field of about 1.5 T. The origin of this slope change is
not clear and needs further exploration. It should be noted that a similar
field induced shift to higher temperature was qualitatively observed in the
peak of specific heat in Ru1212Gd and was interpreted as a possible
signature of p-wave superconductivity\cite{5} However, since no heat
capacity data between 0 and 4 T have been reported a more detailled
comparison with the present resistivity data is not possible and it is not
clear if both phenomena are of the same physical origin.

The key issue to understand the unusual field dependence of $\rho (H)$ is
the magnetic structure involving antiferromagnetic ordering with a weak
ferromagnetic component. The strong field dependence of the intragrain $T_{2}
$ observed in the present data indicates inhomogeneities in the intragrain
structure resulting in a weak link JJA in the superconducting state.
Therefore, phase separation into nanoscale AFM and FM domains may be
considered as a possible scenario.\cite{16} The field $H$ is expected to
change the intra-domain magnetic order as well as the domain structure. The
growth of FM domains with increasing field $H$ will result in enhanced
carrier scattering and a positive magneto-resistance as long as the induced
FM moment does not dominate the AFM order. Only at high field (%
\mbox{$>$}%
5 T) the FM domains determine the transport properties and, due to reduced
carrier scattering, the magneto-resistance should drop again. Percolative
effects could play an essential role.

The physical origin of the observed increase of the resistivity maximum
temperature, $T_{m}$, with the applied field is still an open question. If $%
T_{m}$ is considered as the onset of superconductivity the phenomenon must
be related to the magnetic microstructure and its correlation to the
superconducting state. It is interesting that the field effect on $T_{m}$ is
only observed above 0.4 T where is has been suggested that the principal
axis of alignment of the AFM moments changes from c-axis to in-plane.\cite{3}
This observation calls for a more detailled investigation.

\section{Conclusions}

We have shown that the intragrain superconducting transition temperature of
RuSr$_{2}$EuCu$_{2}$O$_{8}$ decreases steeply as a function of an external
magnetic field. This effect is explained by assuming that the intragrain
superconductivity is due to a phase-lock transition of a nanoscale Josephson
junction array. The resistivity in the ferromagnetically ordered phase and
the temperature of the resistivity maximum both increase as function of
magnetic field above 0.4 T. The positive magnetoresistance and the Josephson
junction nature of the intragrain superconductivity could be explained by a
phase separation into antiferromagnetic and ferromagnetic domains below the
magnetic transition temperature.


\acknowledgments
This work was supported in part by NSF Grant No. DMR-9804325, MRSEC/NSF
Grant No. DMR-9632667, the T. L. L. Temple Foundation, the John and Rebecca
Moores Endowment, and the State of Texas through TCSUH; and at Lawrence
Berkeley Laboratory by the Director, Office of Energy Research, Office of
Basic Sciences, Division of Materials Sciences of the U. S. Department of
Energy under Contract No. DE-AC0376SF00098.


%
%
\begin{figure}[tbp]
\caption{DC susceptibility of bulk and powder sample of RuSr$_{2}$EuCu$_{2}$O%
$_{8}$\newline
circles: ZFC mode; triangles: FC mode\newline
upper and lower lines: powder sample in FC and ZFC mode\newline
inset: M vs. H at 5 K}
\label{Fig. 1}
\end{figure}

\begin{figure}[tbp]
\caption{AC susceptibility of RuSr$_2$EuCu$_2$O$_8$\newline
open triangles and squares: real and imaginary part of bulk sample\newline
solid line: real part of powder sample}
\label{Fig. 2}
\end{figure}

\begin{figure}[tbp]
\caption{Derivative of resistivity at the superconducting transition\newline
(a) zero magnetic field; (b) H=500 Oe\newline
Part (b) shows the decomposition separating intergrain and intragrain
superconducting transitions.\newline
The temperature T$_2$ where the intragrain resistivity drops to 5 \% is
indicated}
\label{Fig. 3}
\end{figure}

\begin{figure}[tbp]
\caption{AC susceptibility of RuSr$_2$EuCu$_2$O$_8$ in a DC bias field of
1000 Oe.\newline
The intragrain transition temperature, T$_2$, is determined from the
crossing of the two straight lines}
\label{Fig. 4}
\end{figure}

\begin{figure}[tbp]
\caption{Field dependence of T$_2$ from resistivity (open triangles) and AC
susceptibility data (closed circles)\newline
The inset shows the Ginzburg-Landau dependence for a bulk superconductor}
\label{Fig.5}
\end{figure}

\begin{figure}[tbp]
\caption{Magnetoresistance, $\protect\rho(50K)$, (open circles) and
temperature of resistivity maximum, T$_m$, (closed squares) as function of
the magnetic field. The inset shows the correlation between $\protect\rho$
and T$_m$.}
\label{Fig. 6}
\end{figure}

%
%

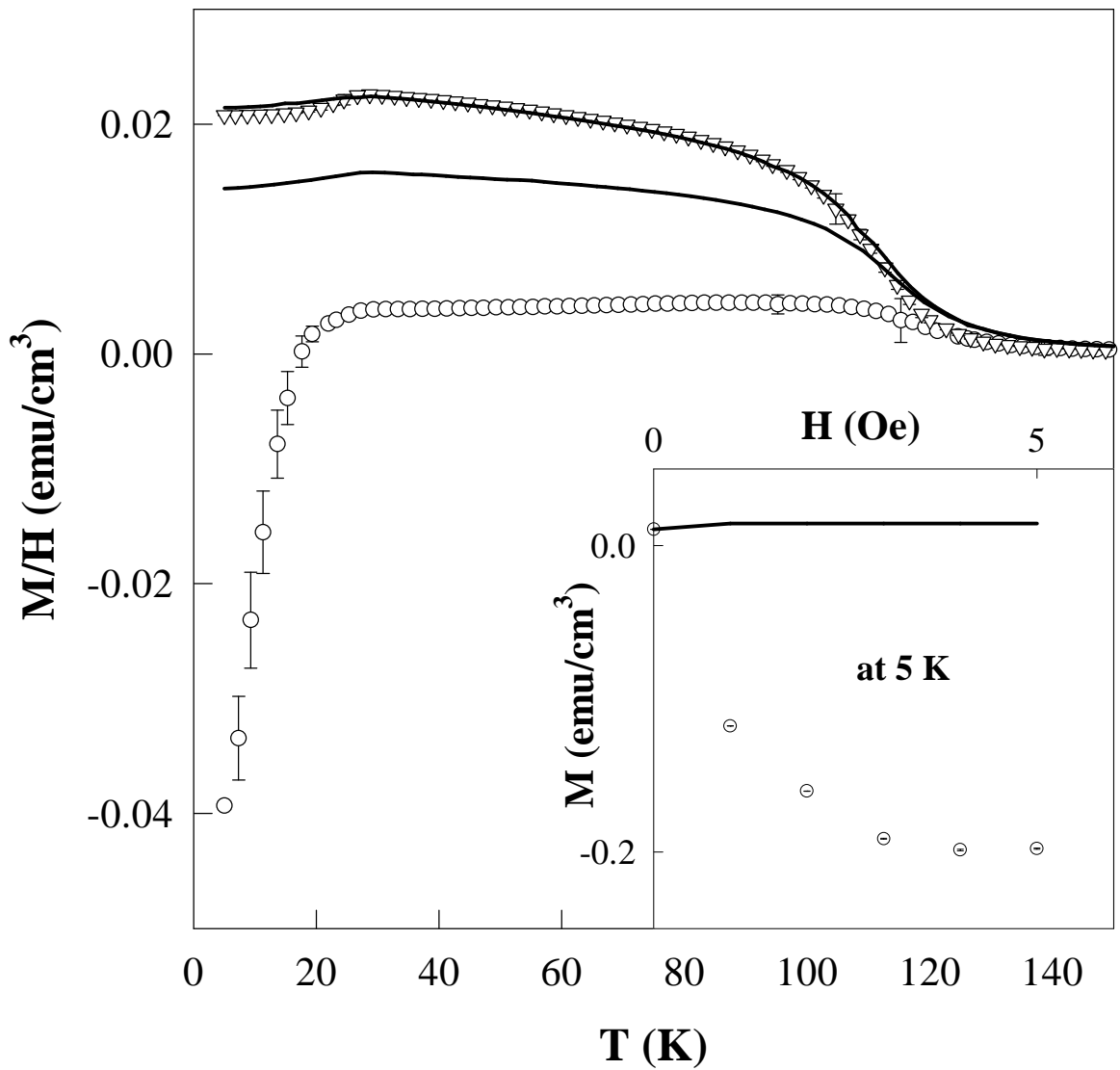
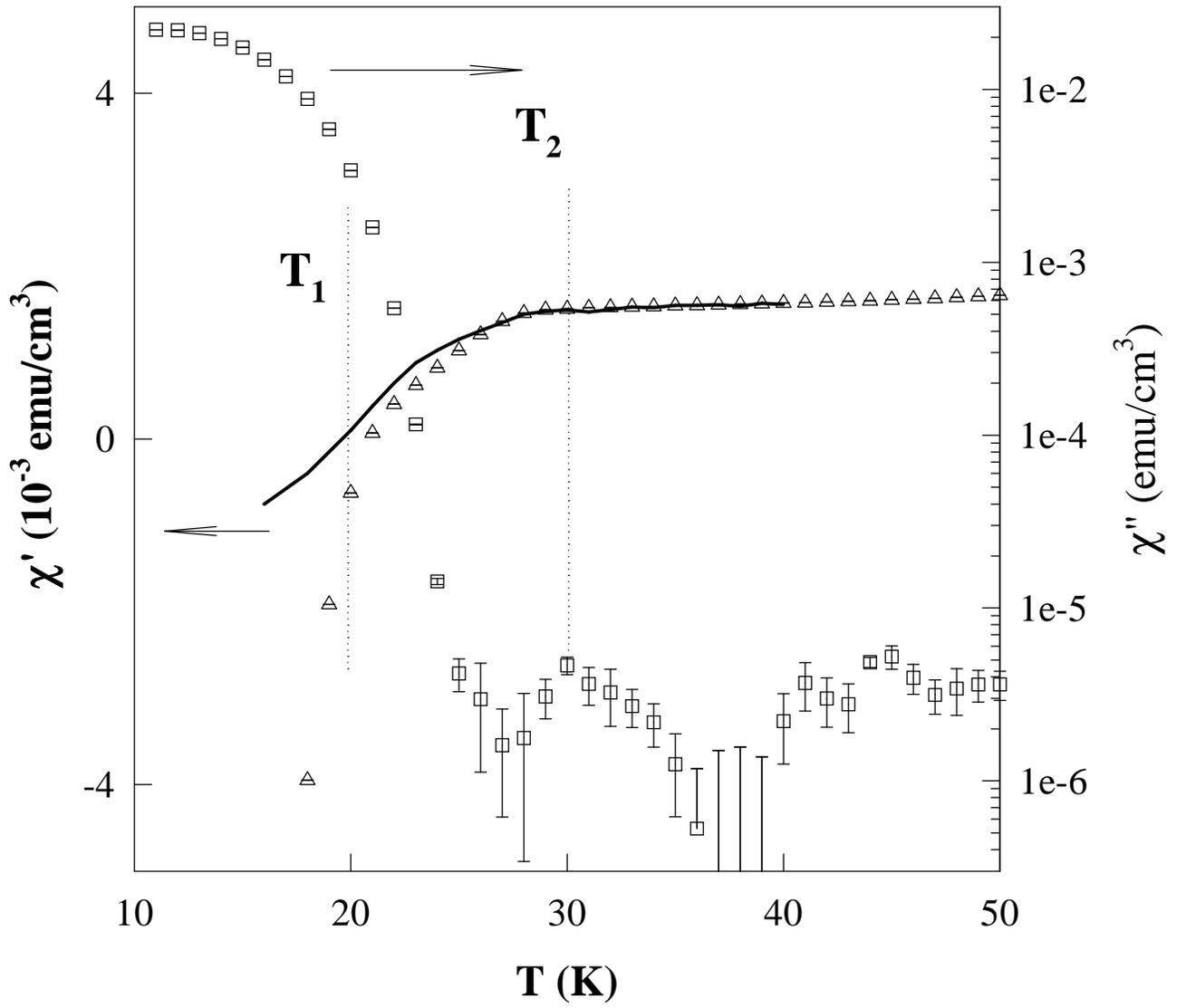
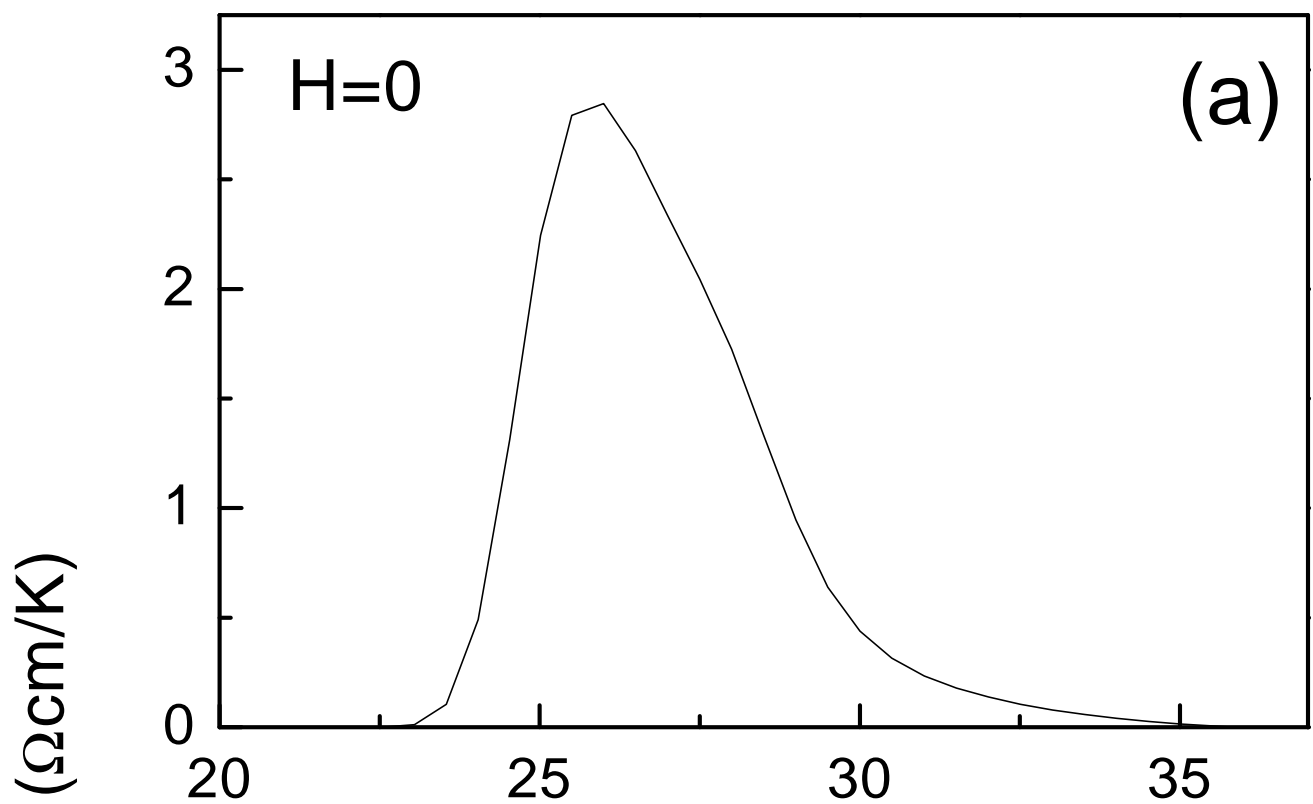
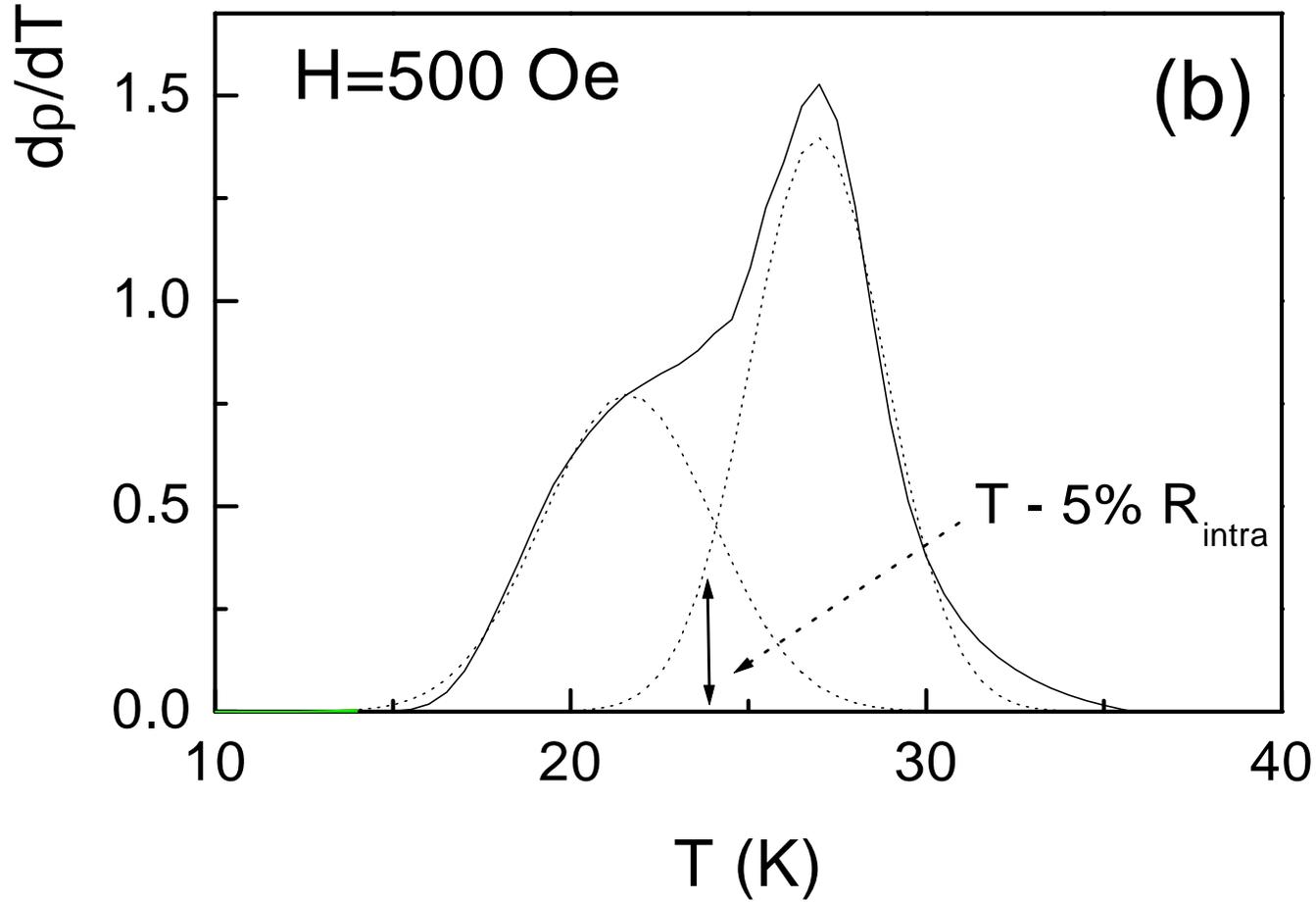
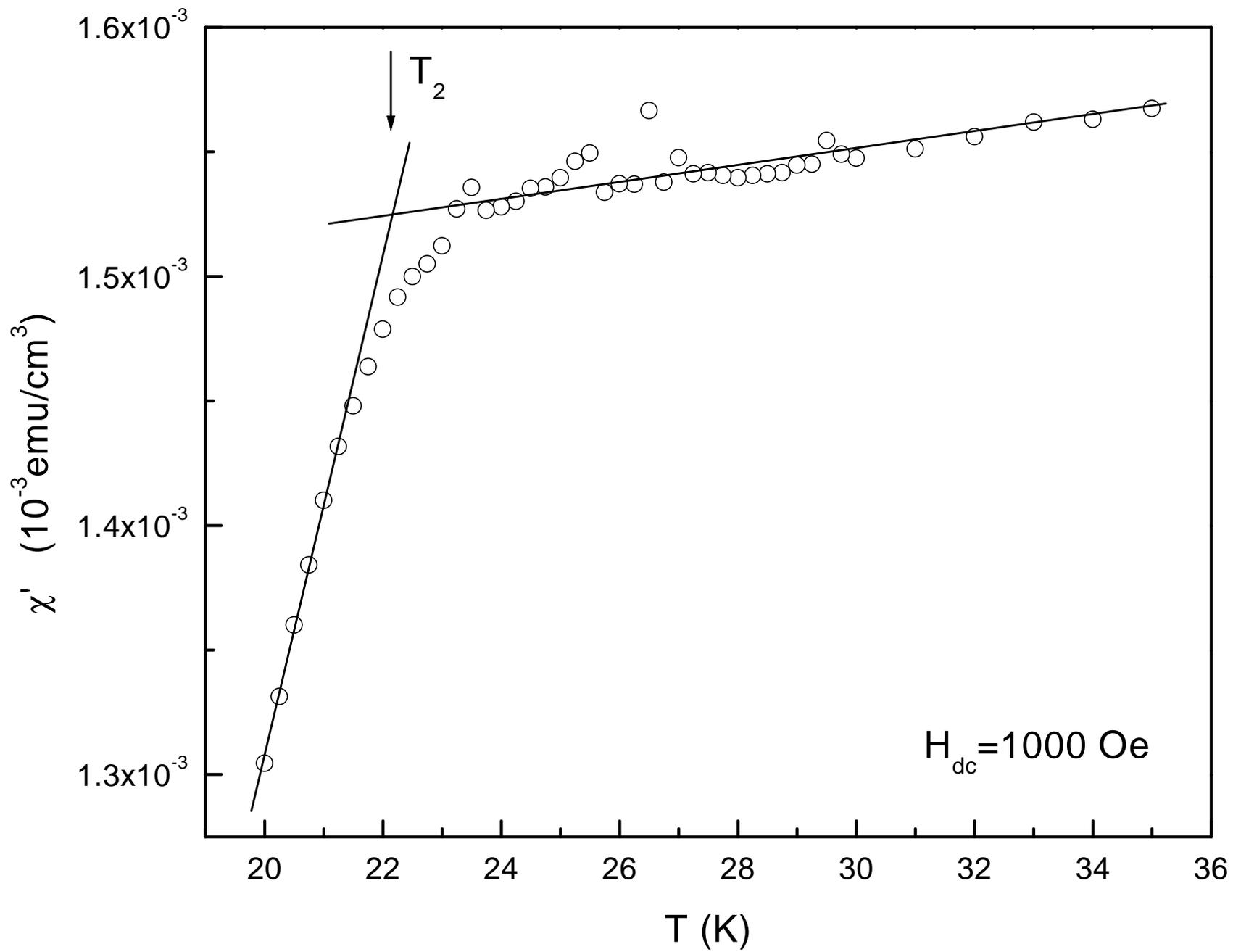
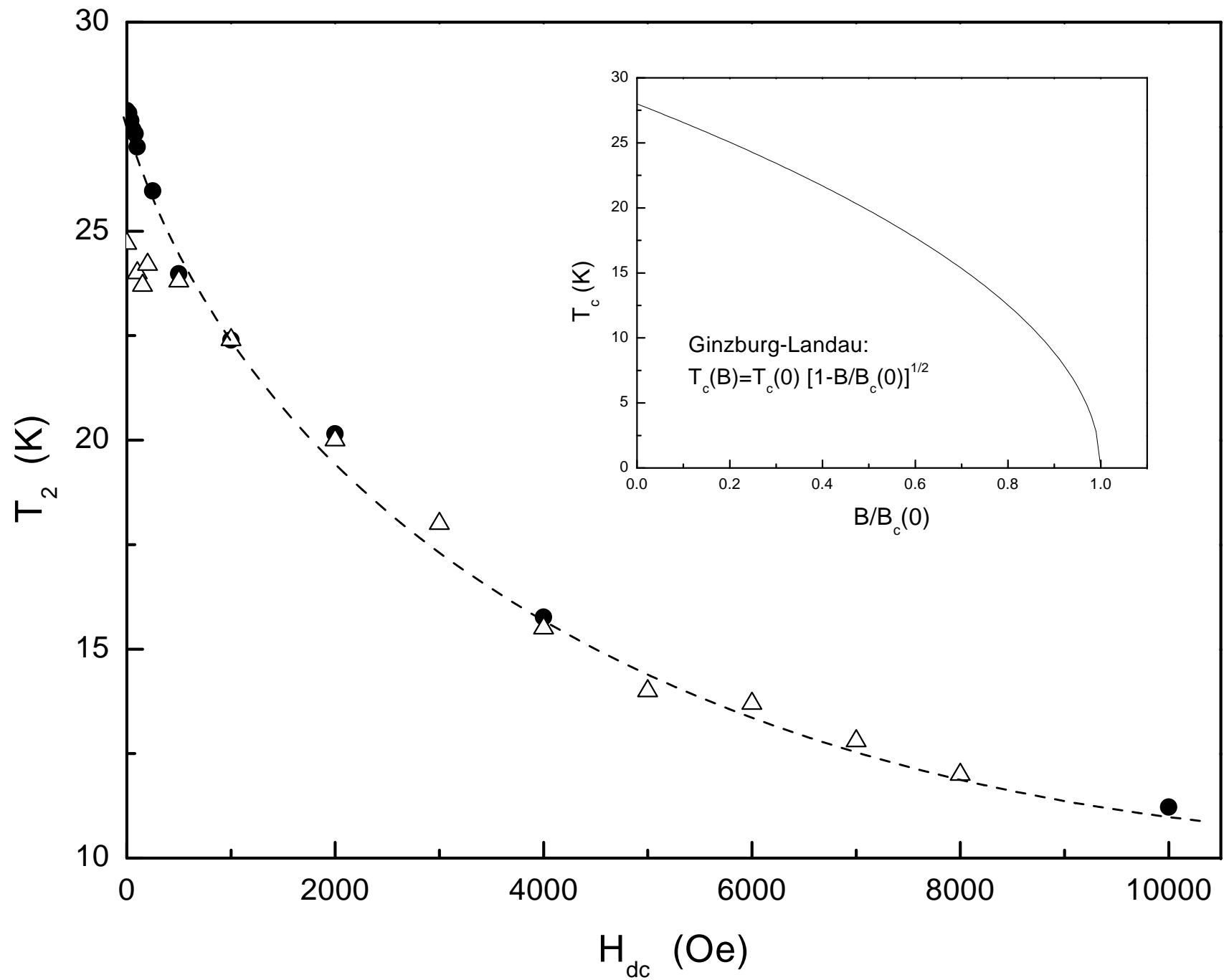
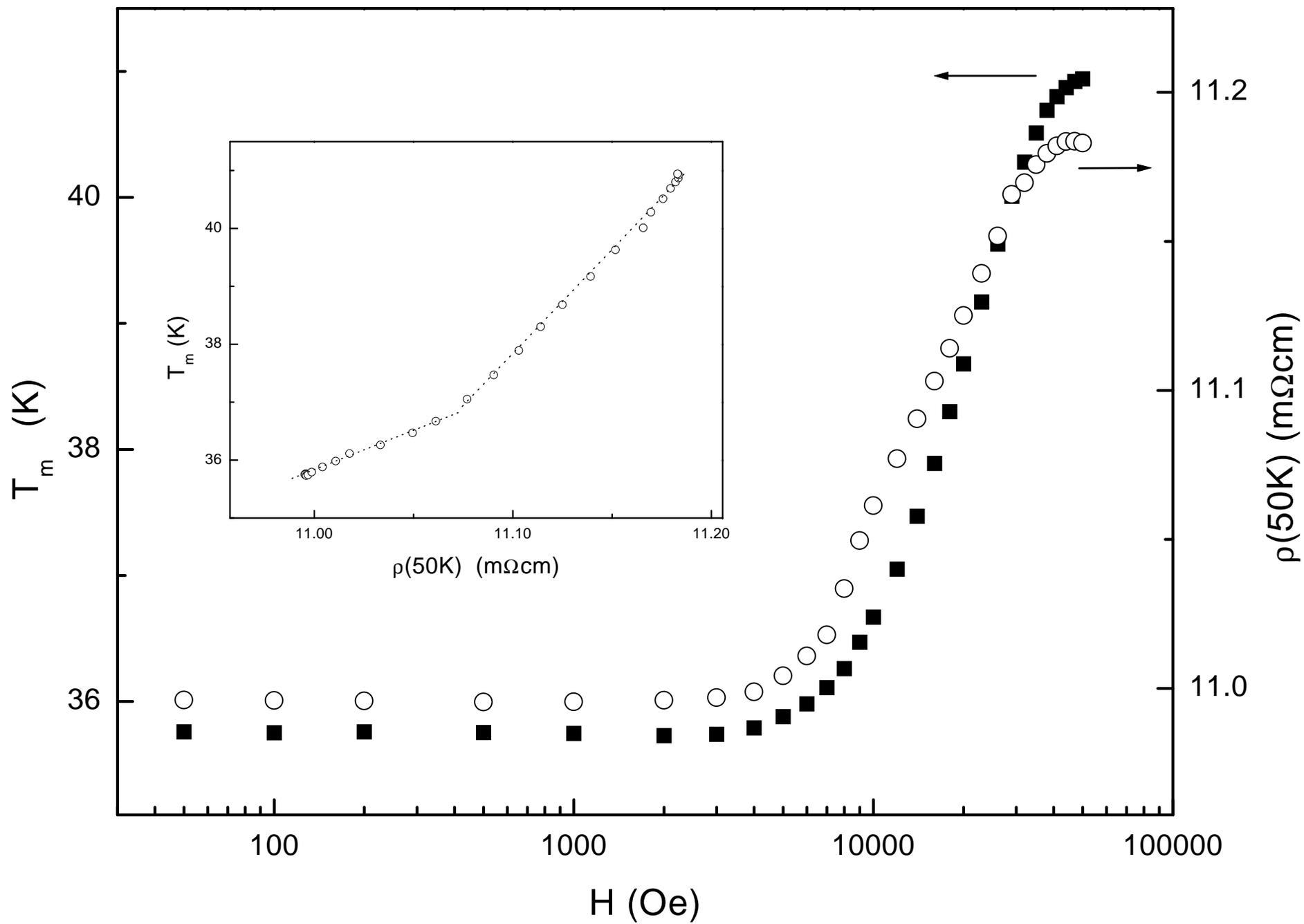
\end{document}